\documentclass[twocolumn,epjc3]{svjour3mod}

\usepackage{hyperref}
\usepackage{graphicx,colordvi}
\usepackage{amsmath,amssymb,slashed,cite}
\usepackage{booktabs,tabulary}
\usepackage{dsfont}
\usepackage{color}

\newcommand{\diff}{\text{d}}
\newcommand{\eV}{\,\text{eV}}
\newcommand{\MeV}{\,\text{MeV}}
\newcommand{\GeV}{\,\text{GeV}}

\renewcommand{\Re}{\text{Re}\,}
\newcommand{\disc}{\text{disc}\,}
\newcommand{\mpi}{M_\pi}
\newcommand{\me}{M_\eta}
\newcommand{\mep}{M_{\eta'}}
\newcommand{\ma}{m_{a_2}}
\newcommand{\eps}{\epsilon}
\newcommand{\eppg}{\eta\to\pi^+\pi^-\gamma}
\newcommand{\epppg}{\eta'\to\pi^+\pi^-\gamma}
\newcommand{\gppe}{\gamma\pi^-\to\pi^-\eta}
\newcommand{\nl}{\notag\\}
\newcommand{\M}{\mathcal{M}}
\newcommand{\Order}{\mathcal{O}}
\newcommand{\F}{\mathcal{F}}
\newcommand{\G}{\mathcal{G}}
\newcommand{\BR}{\mathcal{B}}
\newcommand{\Lagr}{\mathcal{L}}
\newcommand{\beq}{\begin{equation}}
\newcommand{\eeq}{\end{equation}}
\newcommand{\bsp}{\begin{sloppypar}}
\newcommand{\esp}{\end{sloppypar}}

\smartqed  
\RequirePackage{graphicx}

\journalname{Eur. Phys. J. C}
\begin{document}

\title{Anomalous decay and scattering processes of the \boldmath{$\eta$} meson}
\author{Bastian Kubis\thanksref{addr1,addr2}
        \and
        Judith Plenter\thanksref{addr1} 
}
\institute{Helmholtz-Institut f\"ur Strahlen- und Kernphysik,
         Universit\"at Bonn,
         D--53115 Bonn, Germany\label{addr1}
         \and
         Bethe Center for Theoretical Physics,
         Universit\"at Bonn,
         D--53115 Bonn, Germany\label{addr2}
}

\date{}

\maketitle

\begin{abstract}
We amend a recent dispersive analysis of the anomalous $\eta$ decay process 
$\eppg$ by the effects of the $a_2$ tensor meson, the lowest-lying resonance that can
contribute in the $\pi\eta$ system.  While the net effects on the measured decay spectrum are
small, they may be more pronounced for the analogous $\eta'$ decay.
There are nonnegligible consequences for the $\eta$ transition form factor,
which is an important quantity for the hadronic light-by-light scattering 
contribution to the muon's anomalous magnetic moment.
We predict total and differential cross sections, 
as well as a marked forward--backward asymmetry, for the crossed process
$\gppe$ that could be measured in Primakoff reactions in the future. 
\end{abstract}

\section{Introduction}
\label{sec:intro}

The decay $\eppg$ is one of the processes
driven by the chiral anomaly~\cite{WZW_1,WZW_2}.  
The reduced scalar decay amplitude (to be defined below) in the SU(3) chiral limit and at vanishing momenta
is given entirely in terms of the electromagnetic coupling $e$ and the pion decay constant $F_\pi$, 
\beq \label{eq:anomaly}
F_{\eta\pi\pi\gamma} = \frac{e}{4\sqrt{3}\pi^2F_\pi^3} = 5.65\GeV^{-3}.
\eeq
Higher-order corrections to the anomaly can be evaluated in chiral 
perturbation theory~\cite{Bijnens90}, and pion--pion rescattering in the final state
resummed effectively using dispersion theory~\cite{Stollenwerk}.
Besides thus being an interesting decay in its own right to test our understanding
of the interaction of light pseudoscalar mesons with photons, this decay is particularly
noteworthy as a fundamental ingredient in a dispersive analysis of the $\eta$ transition 
form factor $\eta\to\gamma\gamma^*$~\cite{etaTFF}.  This quantity is a crucial input necessary for
the ongoing program to analyze the hadronic light-by-light scattering contribution
to the anomalous magnetic moment of the muon, combining as many pieces of experimental
information as possible in a model-independent fashion~\cite{Bern:lbl,roadmap}.
A similar analysis has also been pursued for the $\pi^0$ transition form factor~\cite{pi0TFF}.

As pointed out in Ref.~\cite{Stollenwerk}, the decays of $\eta$ and $\eta'$ into $\pi^+\pi^-\gamma$
pose a beautiful and simple example to demonstrate the universality of final-state interactions.
Neglecting (tiny) contributions of $F$ and higher partial waves for the pion pair, the authors 
show that the reduced decay amplitude  can be written as
\beq
\F(s,t,u) = P(t) F_\pi^V(t) , \label{eq:PF}
\eeq
where $t=M_{\pi\pi}^2$ is the squared invariant mass of the pion pair, 
$F_\pi^V(t)$ denotes the pion vector form factor as measured in $e^+e^-\to\pi^+\pi^-$, 
and $P(t)$ is a polynomial.  Comparison to experimental data obtained by the WASA-at-COSY~\cite{WASA}
and KLOE~\cite{KLOE} Collaborations demonstrated that within  experimental accuracy,
the polynomial can be assumed to be linear, $P(t)=A(1+\alpha t)$, with~\cite{KLOE}
\beq
\alpha = (1.32 \pm 0.13)\GeV^{-2}. \label{eq:KLOE-alpha}
\eeq

\bsp
This result gives rise to several interesting questions.  Obviously, Eq.~\eqref{eq:PF} is only
an approximation, tested successfully in the physical decay region, $4\mpi^2\leq t \leq \me^2$.  
The universality of final-state interactions expressed therein is only valid in the region
of \emph{elastic} pion--pion rescattering, which is phenomenologically 
a good approximation up to roughly $t \approx 1\GeV^2$.
From generic considerations about the asymptotic behavior of the decay amplitude, one would 
rather expect $P(t)$ to become constant for large $t$, such that the decay amplitude falls 
like $1/t$ similar to the asymptotic behavior of $F_\pi^V(t)$.  
The continuation beyond the physical regime is interesting in particular with regard to 
the application within a dispersive integral to obtain the $\eta$ transition form factor~\cite{etaTFF},
as in principle that integral covers all energies.
\esp

The present article is built on the following observation. 
If we continue the amplitude~\eqref{eq:PF} naively to \emph{negative} $t$, we ought to 
observe a zero at or near $t = - 1/\alpha \approx -0.76\GeV^2$.  Such a kinematical regime is indeed accessible:
in the crossed reaction $\gamma\pi^- \to \pi^-\eta$, which could be measured in a Primakoff-type
reaction, i.e., the scattering of a charged pion in the strong Coulomb field of a heavy
nucleus, producing an additional $\eta$.
Such a Primakoff program is currently pursued by the COMPASS Collaboration (see, e.g., 
Ref.~\cite{KaiserFriedrich} for an overview), using a 
$190\GeV$ $\pi^-$ beam and cutting on very small momentum transfers in order
to isolate the photon-exchange mechanism from diffractive background.  
In this way, COMPASS can investigate $\gamma\pi^-$ reactions to various final states, 
in particular Compton scattering in order to extract the charged-pion polarizabilities~\cite{Gasser:pol,COMPASS:pol},
$\pi^-\pi^0$ to investigate the chiral anomaly~\cite{Nagel,g3pi}, or three pions testing 
chiral predictions~\cite{Kaiser:g4pi,COMPASS:g4pi}.  
In this paper, we want to provide the theoretical motivation to also measure the final state $\pi^-\eta$,
as well as a prediction for the cross sections that are to be expected.

\bsp
For this purpose, beyond using crossing symmetry, we need to amend the amplitude~\eqref{eq:PF}
for the following reason.  The assumption underlying Eq.~\eqref{eq:PF} is the neglect of so-called 
left-hand cuts: the two pions undergoing final-state interactions are assumed to
originate from a point source, such that the amplitude is of form factor type, 
and any interaction (resonant or nonresonant) in the $\pi\eta$ channel is neglected.  
This approximation can be justified
at low energies by appealing to chiral perturbation theory: the $\pi\eta$
$P$-wave  is chirally suppressed (as well as all higher partial waves)~\cite{BKMetapi,etaprimeCusp},
an imaginary part only appears at three-loop order, any phase shift is therefore expected to 
be very small. Furthermore, the $\pi\eta$
$P$-wave has exotic quantum numbers $J^{PC} = 1^{-+}$, and the search for possible resonances 
in this channel is not fully conclusive so far~\cite{COMPASS:pieta,Schott}.
The first well-established resonance that is therefore going to be important in the process
$\gamma\pi\to\pi\eta$ is the $D$-wave tensor meson $a_2(1320)$.  
To investigate its influence is important for several reasons:
\begin{itemize}
\item its inclusion will demonstrate to what extent the feature expected from Eq.~\eqref{eq:PF},
a zero (or at least a pronounced minimum) in certain differential cross sections, 
can survive in a more complete description of the amplitude;
\item it will provide a characteristic breakdown scale in the $\pi\eta$ invariant mass squared
$s = M_{\pi\eta}^2$, above which $\pi\eta$ resonances dominate the cross section;
\item finally, we can use the $a_2$ as the likely most important left-hand-cut structure for the
decay $\eppg$, to study to what extent it affects the decay amplitude, and whether
its effect is consistent with the experimental decay data available.
\end{itemize}

The outline of this article is as follows.  
In Sect.~\ref{sec:eppg}, we recapitulate the dispersive representation of the $\eppg$ decay amplitude
of Ref.~\cite{Stollenwerk}, before calculating contributions of the $a_2$ tensor meson first at tree level, 
then including pion--pion rescattering effects dispersively.  
Section~\ref{sec:decay} compares the resulting observables to the measured $\eppg$ decay spectrum
and briefly discusses the possible impact on the $\eta$ transition form factor.
In Sect.~\ref{sec:scattering}, we give our predictions for the crossed process $\gppe$, discussing
total and differential cross sections, the leading partial waves, as well as the resulting pronounced
forward--backward asymmetry.  We close with a summary.
A brief discussion of the related decay $\epppg$ is relegated to an appendix.
\esp

\section{$\eta\to\pi\pi\gamma$ with left-hand cuts}\label{sec:eppg}

\subsection{Amplitude, kinematics}

We write the decay amplitude for the process
\beq
\eta(q) \to \pi^+(p_1)\pi^-(p_2)\gamma(k)
\eeq
in terms of a scalar function $\F(s,t,u)$ according to
\beq
\M(s,t,u) = i \eps_{\mu\nu\alpha\beta} \eps^\mu(k)p_1^\nu p_2^\alpha q^\beta \F(s,t,u),
\eeq
with the Mandelstam variables given as $s=(q-p_1)^2$, $t=(p_1+p_2)^2$, and $u=(q-p_2)^2$. 
$\F(s,t,u)$ in the chiral limit fulfills the low-energy theorem $\F(0,0,0)=F_{\eta\pi\pi\gamma}$.
The cosine of the $t$-channel center-of-mass angle is given by
\beq
\cos\theta_t = z_t = \frac{s-u}{\sigma_t(\me^2-t)} , \quad \sigma_x = \sqrt{1-\frac{4\mpi^2}{x}}.
\eeq
The $t$-channel partial-wave expansion is of the form
\beq \label{eq:t-pwa}
\F(s,t,u) = \sum_{\text{odd}~l} P'_l(z_t) f_l(t) ,
\eeq
where $P'_l(z_t)$ denote the derivatives of the standard Legendre polynomials.
Due to the strong suppression of $F$ and higher partial waves at low energies, 
we will almost exclusively be concerned with the $P$-wave, which is obtained by angular projection
according to
\beq
f_1(t) = \frac{3}{4}\int_{-1}^1\diff z_t\big(1-z_t^2\big)\F(s,t,u).
\eeq

The differential decay rate with respect to the pion--pion invariant mass squared is given by
\begin{align} \label{eq:dGdt}
\frac{\diff\Gamma}{\diff t} &= \Gamma_0(t) \times
\frac{3}{4}\int_{-1}^1 \diff z_t \big(1-z_t^2\big)|\F(s,t,u)|^2 \nl
&=  \Gamma_0(t) \times \big( |f_1(t)|^2 + \ldots \big) , \nl
\Gamma_0(t) &= \frac{t\sigma_t^3(\me^2-t)^3}{12(8\pi\me)^3} ,
\end{align}
where the ellipsis in the second line represents neglected higher partial waves.

\begin{figure}
\centering
\includegraphics[width=0.55\linewidth]{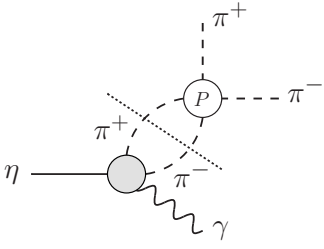}
\caption{Graphical illustration of the discontinuity equation~\eqref{eq:disc}. 
The gray circle denotes the $t$-channel $P$-wave projection of the $\eppg$
decay amplitude, whereas the white circle stands for 
the $P$-wave pion--pion scattering amplitude.}\label{fig:disc}
\end{figure}

\bsp
In the absence of left-hand cuts and ignoring inelasticities, 
the $P$-wave should obey the following representation~\cite{Stollenwerk}:
\beq
f_1(t) = P(t) \Omega(t) , ~
\Omega(t) = \exp\bigg\{\frac{t}{\pi}\int_{4\mpi^2}^\infty\diff x\frac{\delta(x)}{x(x-t)}\bigg\} ,
\label{eq:Omnes}
\eeq
where $\Omega(t)$ is the Omn\`es function~\cite{Omnes} given in terms of the pion--pion $P$-wave
phase shift $\delta(t) \equiv \delta_1^1(t)$, and $P(t)$ is a polynomial.  
The representation~\eqref{eq:Omnes} is a solution to the discontinuity relation
\beq
\disc f_1(t) = 2 i f_1(t) \sin\delta(t) e^{-i\delta(t)} \theta\big(t-4\mpi^2\big) \label{eq:disc}
\eeq
as obtained from elastic pion--pion rescattering; see Fig.~\ref{fig:disc}. 
It obviously
fulfills Watson's final-state interaction theorem~\cite{Watson}: the phase of $f_1(t)$ agrees with
the elastic scattering phase $\delta(t)$.
In the following, we will take $\delta(t)$ from the representation given in Ref.~\cite{Madrid}.
As already pointed out in the introduction, comparison with data~\cite{WASA,KLOE} 
suggested that the polynomial $P(t)$ is linear in the decay region,
\beq \label{eq:P}
P(t) = A(1+\alpha_\Omega t),
\eeq
to very good accuracy.  In fact, in Ref.~\cite{Stollenwerk}, the Omn\`es function was replaced by
the pion vector form factor $F_\pi^V(t)$, which is a phenomenologically 
attractive representation insofar as the latter
is itself directly experimentally observable.  Both representations are equivalent modulo a moderate
shift in the parameter $\alpha \to \alpha_\Omega$ due to the observation that the form factor is in turn proportional
to the Omn\`es function up to a linear polynomial below $1\GeV$, with a slope of the order of 
$0.1\GeV^{-2}$~\cite{etaTFF}.  
\esp

\subsection{Tree-level contribution of the $a_2(1320)$}
\begin{figure}
\centering
\includegraphics[width=\linewidth]{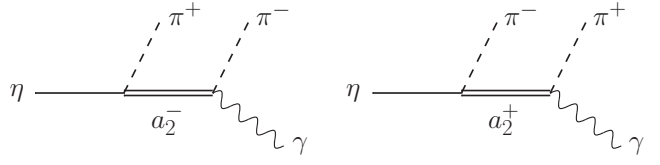}
\caption{Tree-level contributions of the $a_2(1320)$ resonance to $\eppg$ in the 
$s$- (left) and $u$-channel (right).}\label{fig:a2-tree}
\end{figure}

\bsp
We begin by calculating the tree-level contribution of the $a_2$ tensor meson to 
the amplitude $\eppg$ as shown in Fig.~\ref{fig:a2-tree}.
For the formalism of coupling tensor mesons
to Goldstone bosons, we follow Ref.~\cite{EckerZauner}.  The single necessary interaction
term required to describe the decay of a tensor meson into two pseudo\-scalars is given by
\beq
\Lagr_{TPP} = g_T \langle T_{\mu\nu} \{u^\mu,u^\nu\}\rangle , \label{eq:Lagr1}
\eeq
where $\langle.\rangle$ denotes the trace in flavor space.
For simplicity, we only display the nonstrange SU(2) part of the
tensor field relevant to our calculation explicitly,
\beq
T_{\mu\nu} = \frac{1}{\sqrt{2}}\left( \begin{array}{cc} a_2^0 & \sqrt{2}a_2^+ \\ \sqrt{2}a_2^- & - a_2^0 \end{array}\right)_{\mu\nu} + \ldots .
\eeq
The Goldstone bosons are encoded in the field $u_\mu = i (u^\dagger\partial_\mu u- u \partial_\mu u^\dagger )$
(neglecting external currents), where
\beq
u = \exp \Big(\frac{i\phi}{2F_\pi}\Big) , \quad
\phi = \left( \begin{array}{cc} \pi^0 + \frac{\eta}{\sqrt{3}} & \sqrt{2}\pi^+ \\ \sqrt{2}\pi^- & - \pi^0+\frac{\eta}{\sqrt{3}} \end{array}\right) + \ldots . \label{eq:uphi}
\eeq
From Eq.~\eqref{eq:Lagr1}, we can calculate the decay width for $a_2\to\pi\eta$, employing the 
polarization sum of the $a_2$~\cite{EckerZauner}
\begin{align}
\sum_{\text{pol}} \eps_{\mu\nu}(l)\eps_{\rho\sigma}^*(l) &= P_{\mu\nu,\rho\sigma}(l) ,\nl
P_{\mu\nu,\rho\sigma}(l) &= \frac{1}{2}\big(P_{\mu\rho}P_{\nu\sigma}+P_{\nu\rho}P_{\mu\sigma}\big)-\frac{1}{3}P_{\mu\nu}P_{\rho\sigma} ,\nl
P_{\mu\nu} &= g_{\mu\nu}-\frac{l_\mu l_\nu}{\ma^2} ,
\end{align}
and find
\beq \label{eq:TPPwidth}
\Gamma(a_2\to\pi\eta) = \frac{g_T^2}{180\pi F_\pi^4}\frac{\lambda^{5/2}(\ma^2,\mpi^2,\me^2)}{\ma^7},
\eeq
where $\lambda(a,b,c)=a^2+b^2+c^2-2(ab+ac+bc)$ denotes the usual K\"all\'en function.
Equation~\eqref{eq:TPPwidth},
with the total width $\Gamma_{a_2}=(107\pm5)\MeV$
and the branching fraction $\BR(a_2\to\pi\eta) = (14.5\pm1.2) \%$~\cite{PDG}, 
leads to the coupling strength
\beq
|g_T| = (28.1\pm1.4)\MeV ,
\eeq
in perfect agreement with the number obtained in Ref.~\cite{EckerZauner} from the decay $f_2\to\pi\pi$
(compare also Ref.~\cite{Hoferichter:ggpipi}),
thus confirming SU(3) symmetry in this channel.
\esp

The coupling of the $a_2$ to pion and photon can be deduced from a Lagrangian 
(compare Refs.~\cite{Giacosa,Moussallam:ggpipi})
\beq \label{eq:Lagr2}
\Lagr_{TP\gamma} = -\frac{i\,c_T}{2}\eps_{\mu\nu\alpha\beta} 
\langle T^{\alpha\lambda}[f_+^{\mu\nu},\partial^\beta u_\lambda]\rangle ,
\eeq
where $f_+^{\mu\nu}= u F^{\mu\nu}u^\dagger+u^\dagger F^{\mu\nu}u$ (omitting axial vector fields), 
$F^{\mu\nu} = e \mathcal{Q} (\partial^\mu A^\nu-\partial^\nu A^\mu)$ is the electromagnetic
field strength tensor, $\mathcal{Q} = \text{diag}(2/3,-1/3,\ldots)$ the quark charge matrix,
and we have neglected additional currents.  Equation~\eqref{eq:Lagr2} leads to the radiative 
decay width
\beq
\Gamma(a_2\to\pi\gamma) = \frac{e^2c_T^2}{160\pi F_\pi^2}\frac{(\ma^2-\mpi^2)^5}{\ma^5},
\eeq
which, compared to $\BR(a_2\to\pi\gamma)=(2.68\pm0.31)\times10^{-3}$, leads to 
\beq
|c_T| = (0.060\pm0.004)\GeV^{-1} .
\eeq

If we combine the Lagrangians~\eqref{eq:Lagr1} and~\eqref{eq:Lagr2} with the tensor propagator
$i P_{\mu\nu,\rho\sigma}(l)/(\ma^2-l^2)$, we can calculate the $a_2$-exchange contribution 
$\F_{a_2}(s,t,u)$ to 
$\eppg$.  We find
\begin{align} \label{eq:Fa2}
\F_{a_2}(s,t,u) &= \G(s,t,u)+\G(u,t,s), \nl
\G(s,t,u) &= \frac{4ec_Tg_T}{\sqrt{3}F_\pi^3}\frac{1}{\ma^2-s}
\nl
\times \bigg[t&-u+\me^2-\mpi^2-\frac{(s+\mpi^2)(\me^2-\mpi^2)}{\ma^2}\bigg] ,
\end{align}
which is completely fixed by experimental information up to an overall sign.

\bsp
A few remarks are in order concerning Eq.~\eqref{eq:Fa2}.
First, we can also perform an $s$-channel partial-wave expansion 
according to
\begin{align}\label{eq:s-pwa}
\F(s,t,u) &= \sum_{l} P'_l(z_s) g_l(s) , \nl
\cos\theta_s &= z_s = \frac{s(t-u)-\mpi^2(\me^2-\mpi^2)}{(s-\mpi^2)\lambda^{1/2}(s,\me^2,\mpi^2)},
\end{align}
which is the natural partial-wave expansion for $\gppe$ in terms of the 
scattering angle $\theta_s$.  The partial-wave expansion of the $s$-channel $a_2$ exchange amplitude
$\G(s,t,u)$ then reads
\begin{align} \label{eq:a2P+D}
\G(s,t,u) &= g_1^{a_2}(s) + 3z_s g_2^{a_2}(s) ,\nl
g_1^{a_2}(s) &= \frac{4ec_Tg_T}{\sqrt{3}F_\pi^3}
\frac{(s+\mpi^2)(\me^2-\mpi^2)}{s\,\ma^2},\nl
3z_s g_2^{a_2}(s) &= \frac{4ec_Tg_T}{\sqrt{3}F_\pi^3}\frac{1}{\ma^2-s}
\bigg[t-u-\frac{\mpi^2(\me^2\!-\!\mpi^2)}{s}\bigg].
\end{align}
Phrased differently, $\G(s,t,u)$ contains a nonresonant $P$-wave contribution (which has no $a_2$ propagator) 
in addition to the expected resonant
$D$-wave.  This is a well-known problem of higher-spin propagators; see e.g.\ the discussion in Ref.~\cite{EckerZauner}.
We cannot easily subtract the $P$-wave and use the $D$-wave alone, as Eq.~\eqref{eq:a2P+D} shows that 
both partial waves individually display an artificial pole $\propto 1/s$, which is not present in the 
full amplitude~\eqref{eq:Fa2}.  While a pole at $s=0$ is not kinematically accessible in either of the two processes
we consider in this article, it precludes a dispersive reconstruction of $t$-channel rescattering as discussed in the 
following section.  We therefore retain the $P$-wave part in Eq.~\eqref{eq:a2P+D}; its effect turns out to be numerically
small. 

Second, we fix the sign of $c_Tg_T$ in the following way. As pointed out in Ref.~\cite{Stollenwerk}, 
the vector-meson contributions to $\eppg$ determined in Ref.~\cite{Bijnens90} can be rewritten, 
using the limit of a large number of colors (i.e., neglecting loop effects)
and expanding the $\rho$ propagators to leading order in the spirit of resonance saturation
of chiral low-energy constants, as
\begin{align}
\F(s,t,u) &= F_{\eta\pi\pi\gamma} \bigg[ 1+ \frac{3t}{2m_\rho^2} + \Order\big(m_\rho^{-4}\big) \bigg] \nl
&= F_{\eta\pi\pi\gamma} \bigg[ 1+ \frac{t}{2m_\rho^2} \bigg]  \Omega(t) + \Order\big(m_\rho^{-4}\big) , \label{eq:VMD}
\end{align}
where we have used the approximation
\beq
\Omega(t) \approx \frac{m_\rho^2}{m_\rho^2-t} = 1+\frac{t}{m_\rho^2}+ \Order\big(m_\rho^{-4}\big) .
\eeq
In other words, Eq.~\eqref{eq:VMD} predicts $\alpha_\Omega^\rho \approx 1/(2m_\rho^2) = 0.84\GeV^{-2}$, 
a little more than half of the phenomenological value $\alpha_\Omega \approx 1.52\GeV^{-2}$ 
(when using Eq.~\eqref{eq:Omnes} for the definition of $\alpha_\Omega$ and not the pion vector form factor
for $\alpha$).
We can now similarly expand Eq.~\eqref{eq:Fa2} to leading order in inverse powers of $\ma^2$.  If we neglect
the induced quark mass renormalization of the anomaly (proportional to $\mpi^2$, $\me^2$), we
find the following estimate for the $a_2$ contribution to $\alpha_\Omega$:
\beq
\alpha_\Omega^{a_2} = \frac{48\pi^2 c_T g_T}{\ma^2} = \pm (0.46\pm0.04)\GeV^{-2}. \label{eq:alpha-a2}
\eeq
We shall see below that the true effect when including the $a_2$ in a new extraction of the
slope parameter $\alpha_\Omega$ from data is significantly smaller, mainly due to curvature effects
in the induced amplitude.  Still, while effects in particular of excited $\rho'$ resonances
can be nonnegligible, we take the discrepancy between the $\rho$-induced slope
and the experimentally determined value as an indication that the sign of the $a_2$ contribution
ought to be positive,
\beq \label{eq:sign-a2}
c_Tg_T = + |c_Tg_T|.
\eeq
The point of view of chiral perturbation theory allows us to further substantiate this choice.
If we add the amplitude~\eqref{eq:Fa2},
expanded to leading order in $1/\ma^2$ and with the sign as in Eq.~\eqref{eq:sign-a2}, 
as a further resonance saturation contribution to 
the one-loop representation of Ref.~\cite{Bijnens90}, the partial width $\Gamma(\eppg)$ 
increases by about $7\eV$, bringing the original prediction of $47\eV$ into even better 
agreement with the experimental number $(55\pm2)\eV$~\cite{PDG}.
We will therefore work on from this hypothesis, and we give further hints below that data indeed suggests
this to be the more likely solution.

As a final remark, we will later insert a nonvanishing, energy-dependent width in the $a_2$ propagator 
in Eq.~\eqref{eq:Fa2} by hand,
\beq
\frac{1}{\ma^2-s} \longrightarrow \frac{1}{\ma^2-s - i \ma \Gamma_{a_2}(s)} ,
\eeq 
using the parametrization~\cite{VES:a2} 
\begin{align} \label{eq:a2width}
\Gamma_{a_2}(s) &= \Gamma_{a_2} \sum_{i=\eta,\rho} p_i \frac{\ma}{\sqrt{s}}\frac{q_i(s)}{q_i(\ma^2)}
\frac{T\big(q_i(s)R\big)}{T\big(q(\ma^2)R\big)}  ,\nl
q_{\eta/\rho}(s) &= \frac{\lambda^{1/2}\big(s,M_{\eta/\rho}^2,\mpi^2\big)}{2\sqrt{s}}, \quad
T(x) = \frac{x^4}{9+3x^2+x^4} ,
\end{align}
which explicitly takes into account the $a_2$ decays into final states $\pi\eta$ and $\pi\rho$
with relative branching fractions $p_\eta=0.17$, $p_\rho=0.83$, using the barrier factor
$R=5.2\GeV^{-1}$.  
The $a_2$ is sufficiently far from the $\pi\rho$ ``threshold'' that it seems a justifiable  
approximation to treat the $\rho$ as a stable particle in this case.
In contrast to using a constant width, this parametrization provides the correct
threshold behavior of the imaginary part, as well as a reasonable phase above the resonance.
\esp

\subsection{Unitarization}

\begin{figure}
\centering
\includegraphics[width=0.6\linewidth]{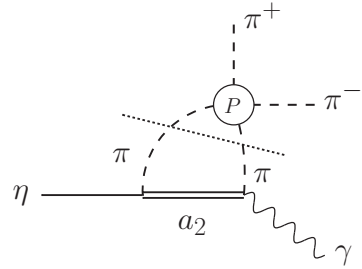}
\caption{$a_2$ contribution with pion--pion $P$-wave rescattering.
The graph is supposed to comprise both possible charge assignments inside the loop,
corresponding to tree-level $s$- and $u$-channel $a_2$ exchange; see Fig.~\ref{fig:a2-tree}.}\label{fig:a2-loop}
\end{figure}

It is obvious that simply adding the tree-level $a_2$ contribution~\eqref{eq:Fa2} to the 
original amplitude~\eqref{eq:Omnes} violates Watson's theorem: we are missing the pion--pion rescattering
on top of the $a_2$-exchange graphs; see Fig.~\ref{fig:a2-loop}.
The full dispersive solution that reinstates the correct phase relation in the $t$-channel $P$-wave is of the form
\begin{align} \label{eq:full}
\F(s,t,u) &= \F(t)+\G(s,t,u)+\G(u,t,s), \nl
\F(t) &= \Omega(t)\bigg\{ A(1+\alpha_\Omega t) \nl &  \qquad \qquad + \frac{t^2}{\pi}\int_{4\mpi^2}^\infty
\frac{\diff x}{x^2} \frac{\sin\delta(x) \hat\G(x)}{|\Omega(x)|(x-t)} \bigg\}, \nl
\hat\G(t) &= \frac{3}{4}\int_{-1}^1\diff z_t\big(1-z_t^2\big) \big[\G(s,t,u)+\G(u,t,s)\big] .
\end{align}
$\hat\G(t)$ is the projection of the $a_2$ exchange graphs onto the $t$-channel $P$-wave.
It is given explicitly by
\begin{align}
\hat\G(t)&= \frac{8ec_Tg_T}{\sqrt{3}F_\pi^3} \bigg\{
\frac{\me^2-\mpi^2}{\ma^2} -1 + 
\frac{1}{\me^2-t}\bigg[\ma^2 \nl
& \quad +2t-3\mpi^2- \frac{(\ma^2+\mpi^2)(\me^2-\mpi^2)}{\ma^2} \bigg] Q(y)   , \nl
Q(y) &= \frac{3}{\sigma_t} \bigg( y + \frac{1-y^2}{2}\log\frac{y+1}{y-1}\bigg) , \nl
y &= \frac{2\ma^2-\me^2-2\mpi^2+t}{\sigma_t(\me^2-t)} .
\end{align}
$\hat\G(t)$ contains a square-root singularity at $t=0$, signaling the onset of the left-hand cut.
As $\hat\G(t)$ approaches a constant for large arguments $t\to\infty$, two subtractions in Eq.~\eqref{eq:full}
are sufficient, as the Omn\`es function behaves asymptotically as 
$\Omega(t) \sim 1/t$ for $\delta(t) \to \pi$.  The number of
subtractions therefore exactly reflects the original form in Eqs.~\eqref{eq:Omnes} and \eqref{eq:P}.

It is easy to see that the full $t$-channel $P$-wave resulting from Eq.~\eqref{eq:full},
\beq
f_1(t) = \F(t) + \hat\G(t), \label{eq:f1}
\eeq
has the correct phase, while $\F(t)$ alone is subject to the inhomogeneous discontinuity relation
\begin{align}
\disc\F(t) &= 2i\big[\F(t)+\hat\G(t)\big]\sin\delta(t)e^{-i\delta(t)}\theta\big(t-4\mpi^2\big) . 
\end{align}
The representation~\eqref{eq:full}, using the inhomogeneity $\hat\G(t)$ as input to the dispersive
integral, preserves unitarity in the $t$-channel in the presence of left-hand cuts, which 
are approximated by resonance (here: $a_2$) contributions.  This is closely related to the methods
used e.g.\ in Ref.~\cite{Moussallam:ggpipi} for $\gamma\gamma\to\pi\pi$, or in Ref.~\cite{Bl4} for
semileptonic $B$-decays.  We cannot easily apply an iterative procedure to determine left-hand cuts from 
right-hand cuts and vice versa, as done e.g.\ in the analysis of the closely related 
Primakoff process $\gamma\pi\to\pi\pi$~\cite{g3pi}, as we do not have independent information on 
$\pi\eta$ scattering phases at our disposal.  

\bsp
Obviously, the $a_2$ $s$- and $u$-channel exchanges will also generate nonvanishing projections onto 
$F$- and higher $t$-channel partial waves.  These partial waves are real as long as we neglect pion--pion 
rescattering effects in those higher waves, which is entirely justified for $\eppg$
(and even for $\epppg$), given the smallness of the corresponding phases; 
compare the discussion in Ref.~\cite{V3pi}.
However, even the real part of the $F$-wave is entirely negligible: while in the chiral power counting,
it is suppressed compared to the $P$-wave by another power of $p^2/\ma^2$, 
we have checked that kinematical prefactors effectively suppress it by more than 3 orders of magnitude
in the physical decay region of $\eppg$, and still by 2 for $\epppg$.
We will therefore discuss the comparison to decay data in the following section still in the approximation
indicated in Eq.~\eqref{eq:dGdt}, using the $P$-wave only.
\esp

\section{Comparison to decay data}\label{sec:decay}

\subsection{$\eppg$ decay spectrum}

In this section, we compare the amplitude constructed in the previous section to the 
data on $\diff\Gamma/\diff t$ as obtained by the KLOE Collaboration~\cite{KLOE}.
The decay distribution was measured with arbitrary normalization, 
which has to be fixed independently from the branching fraction $\BR(\eppg) = (4.22\pm 0.08)\%$,
as well as the total width of the $\eta$~\cite{PDG}.

We first (re)fit the representation~\eqref{eq:Omnes}, \eqref{eq:P}.  We obtain
\beq
\alpha_\Omega = (1.52 \pm 0.06)\GeV^{-2} , \label{eq:alpha-fit1}
\eeq
where the error is only due to the statistical uncertainty in the data and
neglects all the systematic effects discussed in Ref.~\cite{KLOE}. 
The difference in the central value compared to $\alpha$ in Eq.~\eqref{eq:KLOE-alpha}
is due to employing the Omn\`es function instead of the pion
vector form factor, as discussed above.\footnote{In fact, if we construct the Omn\`es function 
from the phase of the pion vector form  factor 
instead of from the $\pi\pi$ $P$-wave phase shift~\cite{Madrid} as in Ref.~\cite{omegaTFF},
the central value of $\alpha_\Omega$ reduces to $1.37\GeV^{-2}$, rather close to Eq.~\eqref{eq:KLOE-alpha}.
We disregard the effects of varying the $\pi\pi$ phase input in the following: they are compensated 
by corresponding shifts in $\alpha_\Omega$ to a very large extent, and they lead to insignificant uncertainties
compared to other error sources.}
The quality of the fit is excellent, with a $\chi^2$ per degree of freedom of $\chi^2/\text{ndof} = 0.94$.  
The subtraction constant $A$ that, in this case, serves as an overall normalization of the amplitude,
is $A= (5.43\pm0.12\mp0.04)\GeV^{-3}$, where the first error is due to the uncertainty in the integrated 
partial width, and the second due to the uncertainty in $\alpha_\Omega$, almost perfectly anticorrelated with the latter.
$A$ thus seems well compatible with $F_{\eta\pi\pi\gamma}$, see Eq.~\eqref{eq:anomaly}.

\begin{figure}
\includegraphics[width=\linewidth,clip]{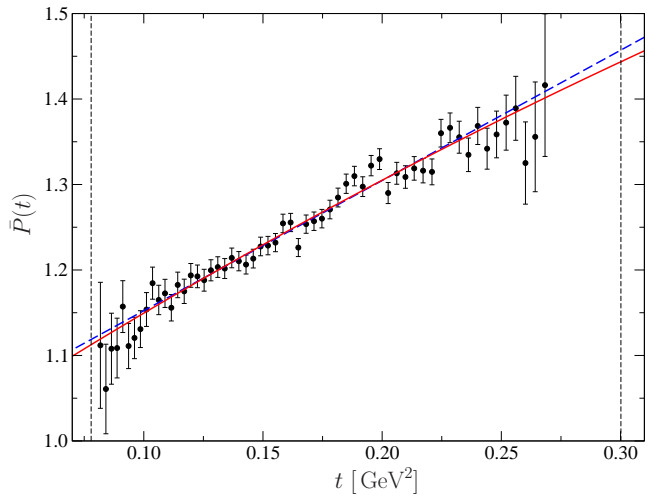}
\caption{Representation of the decay distribution $\eppg$ from Ref.~\cite{KLOE};
see main text for details.  The blue dashed curve shows the linear fit, while the full red
curve includes the effects of $a_2$ exchange in addition.  The vertical dashed lines represent
the limits of phase space at $4\mpi^2$ and $\me^2$.} \label{fig:KLOElinear}
\end{figure}
In Fig.~\ref{fig:KLOElinear}, we plot the following observable, obtained from the data from Ref.~\cite{KLOE}:
\beq \label{eq:defPbar}
\bar P(t) = \sqrt{\frac{1}{\Gamma_0(t)} \frac{\diff\Gamma}{\diff t}} \Big/ \big[ A \Omega(t) \big] ,
\eeq
i.e.\ within the accuracy of the amplitude representation without left-hand cuts, we expect
to find $\bar P(t) = P(t)/P(0) = 1+\alpha_\Omega t$.  As the quality of the fit suggests, the linear curve (blue dashed)
describes the data perfectly.

\bsp
Including the effects of $a_2$ exchange (properly unitarized in the $t$-channel), 
the subtraction constant $\alpha_\Omega$ in Eq.~\eqref{eq:full} has to be refitted to the data.  We obtain
\beq
\alpha_\Omega = (1.42 \pm 0.06)\GeV^{-2} , \label{eq:alpha-fit2}
\eeq
with $\chi^2/\text{ndof} = 0.90$.  
The uncertainty of the $a_2$ coupling constants induces an additional error in $\alpha_\Omega$ 
of $\pm0.01\GeV^{-2}$, which we will neglect in the following.
The resulting fit is also shown in Fig.~\ref{fig:KLOElinear}.
The reduction in the value of $\alpha_\Omega$ compared to Eq.~\eqref{eq:alpha-fit1} may seem surprisingly small,
given the estimate of the $a_2$ contribution to this parameter in Eq.~\eqref{eq:alpha-a2}.  
The reason is the curvature in $\bar P(t)$: 
in fact, the derivative $\bar P'(t)$ (which equals the constant $\alpha_\Omega$ 
in the simple fit)
varies from $\bar P'(4\mpi^2) = 1.69\GeV^{-2}$ to $\bar P'(\me^2) = 1.30\GeV^{-2}$ within the decay phase space; 
outside phase space, we find e.g.\ $\bar P'(1\GeV^2) = 0.46\GeV^{-2}$, and naive continuation to yet higher energies
makes the derivative vanish and change sign around $\sqrt{t}=1.25\GeV$.
It finally diverges at $t=0$ due to the square-root singularity.  
\esp

\subsection{Impact on the $\eta$ transition form factor}

As far as the phenomenological description of the $\eppg$ decay data of Ref.~\cite{KLOE}
is concerned, the two amplitudes, with and without $a_2$ effects included, are clearly equivalent:
they describe the data equally well, and in fact, the two fit curves displayed in Fig.~\ref{fig:KLOElinear}
deviate from each other by less than $1\%$ in the whole decay region.
This is different in the wider kinematic range of the similar decay $\epppg$, 
which we discuss in \ref{app:etaprime}.  While the available data do not yet allow one to prefer
one amplitude over the other in a statistically valid sense, the comparison of the extracted subtraction 
constants $\alpha_\Omega$ and an $\alpha'_\Omega$ defined in an analogous manner  
for $\epppg$ seems to favor somewhat the decay amplitude including the curvature effects
induced by the $a_2$.

However, we have emphasized in the introduction that the decay amplitude $\eppg$ serves
as a crucial input to a dispersive analysis of the $\eta$ transition form factor~\cite{etaTFF}, 
where the dispersion integral extends over a much larger range in energy (in principle, up to infinity).
We therefore may expect to see a somewhat more significant deviation between the two amplitudes 
in there.  

\begin{figure}
\centering
\includegraphics[width=0.55\linewidth]{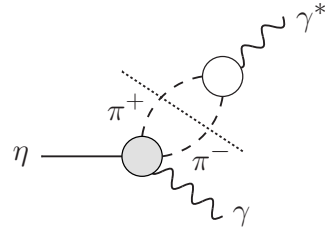}
\caption{Two-pion cut contribution to the isovector part of the (singly virtual) $\eta$ 
transition form factor.  Here, the gray circle denotes the $t$-channel $\eppg$ $P$-wave,
while the white circle is the pion vector form factor.}\label{fig:disc-TFF}
\end{figure}

\bsp
We refer e.g.\ to Ref.~\cite{etaTFF} for all pertinent definitions concerning the singly virtual 
$\eta$ transition form factor $F_{\eta\gamma^*\gamma}(Q^2,0)$, which at small photon virtualities can be 
expanded according to
\beq
\frac{F_{\eta\gamma^*\gamma}(Q^2,0)}{F_{\eta\gamma^*\gamma}(0,0)} = 
1+\Big(b_\eta^{(I=1)}+b_\eta^{(I=0)}\Big) Q^2
+\Order(Q^4).
\eeq
The slope parameter $b_\eta$ is divided into an isovector $I=1$ and an isoscalar $I=0$ piece.
The isoscalar part is small: employing $\omega+\phi$ dominance
together with data input on $\omega,\phi\to\eta\gamma$ yields 
$b_\eta^{(I=0)}\approx -0.022\GeV^{-2}$~\cite{etaTFF}.  The slope is therefore almost entirely
given by the iso\-vector contribution, which in turn is dominated by $\pi^+\pi^-$ intermediate 
states; see Fig.~\ref{fig:disc-TFF}.
The corresponding sum rule can be written as~\cite{etaTFF} 
\beq
b_\eta^{(I=1)} = \frac{e}{96\pi^2 A^\eta_{\gamma\gamma}} \int_{4\mpi^2}^{\Lambda^2} \frac{\diff x}{x}\sigma_x^3 
F_\pi^{V*}(s)f_1(x), 
\eeq
where $F_\pi^V(t)$ is the standard pion vector form factor,
and we have written the dispersion integral with a cutoff $\Lambda^2$ instead of integrating to infinity.
The $\eta\to\gamma\gamma$ amplitude $A^\eta_{\gamma\gamma}$ is obtained from the corresponding partial width by
\beq
A^\eta_{\gamma\gamma} = \sqrt{\frac{64}{\me^3}\Gamma(\eta\to\gamma\gamma)}.
\eeq
\esp

Following Ref.~\cite{etaTFF}, we vary the cutoff in the range $\Lambda^2 = \{\mep^2\ldots 2\GeV^2\}$.
With the decay amplitude~\eqref{eq:Omnes}, \eqref{eq:P}, we find
\begin{align}
b_\eta^{(I=1)} = \big[ & \{2.04\ldots2.22\} \pm 0.04_\alpha \nl & \pm 0.02_\BR \pm 0.01_{F_\pi^V} \big]\GeV^{-2},
\end{align}
where the indicated range follows the range of cutoffs, and the errors are due to uncertainties 
in $\alpha_\Omega$,\footnote{While we have propagated the statistical error on $\alpha_\Omega$ from Eq.~\eqref{eq:alpha-fit1}
only in the rest of this paper, we here use the larger uncertainty $\pm0.13$ due to systematic
effects~\cite{KLOE} in order to be consistent with the analysis in Ref.~\cite{etaTFF}.} 
the branching ratios for $\eppg$ and $\eta\to\gamma\gamma$, 
and the pion vector form factor.  For the latter, we employ the pion vector form factor parametrizations of 
Refs.~\cite{Hanhart:pionFF,omegaTFF} (or approximations thereof).
Using, however, the partial wave $f_1(t)$ as in Eq.~\eqref{eq:f1}, the result reduces to
\begin{align}
b_\eta^{(I=1)} = \big[ & \{1.90\ldots2.04\} \pm 0.04_\alpha \nl & \pm 0.02_\BR \pm 0.01_{F_\pi^V}\pm 0.01_{a_2}\big]\GeV^{-2},
\end{align}
with the additional error due to the uncertainty in the $a_2$ coupling constants.
That is, the slope is reduced by about $7\%$, a bit more than the combined error cited in Ref.~\cite{etaTFF},
for a cutoff $\Lambda^2\approx 1\GeV^2$; this reduction is increased for higher cutoffs (due to the 
increasingly stronger curvature effects).
A more detailed investigation of $a_2$ effects on the $\eta$ (and $\eta'$) transition form factor(s),
beyond the value of the slope at the origin, should still be pursued.

\section{Phenomenology for $\gamma\pi\to\pi\eta$}\label{sec:scattering}

\bsp
In the previous section, we have constructed an $\eppg$ decay amplitude including the leading
left-hand-cut contribution, and have shown that this amplitude describes the available decay data very well.
As this representation includes the lightest resonance that can contribute in the $\pi\eta$ system, 
we are well equipped to now consider the crossed process
\beq
\gamma(k)\pi^-(p_2) \to \pi^-(\bar p_1)\eta(q),
\eeq
which is described by the same amplitude as the decay process in Sect.~\ref{sec:eppg}
with $\bar p_1=-p_1$ (using time-reversal invariance).  
The Mandelstam variables are defined as before, e.g.\ $s=(\bar p_1+q)^2$ denotes
the total energy squared in the center-of-mass system, $t=(p_2-\bar p_1)^2$ is related to the 
pion momentum transfer etc.  In particular, Eq.~\eqref{eq:s-pwa} is the natural partial-wave
expansion in scattering kinematics.

The (polarization-averaged) differential cross section is given by 
\beq
\frac{\diff\sigma}{\diff\Omega} = \frac{(s-\mpi^2)\lambda^{3/2}(s,\mpi^2,\me^2)}{2048\pi^2s^2} \big(1-z_s^2\big)
 |\F(s,t,u)|^2,
\eeq
from which one obtains for the total cross section
\begin{align}
\sigma(s) &= \frac{(s-\mpi^2)\lambda^{3/2}(s,\mpi^2,\me^2)}{1024\pi s^2} \nl
&  \quad \times \int_{-1}^1\diff z_s \big(1-z_s^2\big)
 |\F(s,t,u)|^2 \nl
&= \frac{(s-\mpi^2)\lambda^{3/2}(s,\mpi^2,\me^2)}{768\pi s^2} \nl
& \quad \times \bigg[ |g_1(s)|^2 + \frac{9}{5}|g_2(s)|^2 + \frac{18}{7}|g_3(s)|^2 + \ldots \bigg],
\end{align}
where we have inserted the $s$-channel partial-wave expansion~\eqref{eq:s-pwa} up to $F$-waves in the second step.

\begin{figure}
\centering
\includegraphics[width=0.5\linewidth]{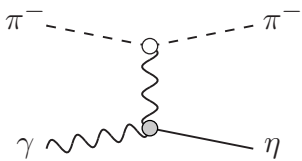}
\caption{Radiative correction to $\gppe$ due to photon-exchange diagram.}\label{fig:photon-exchange}
\end{figure}

As a cautionary side remark, we wish to point out that it has been emphasized
in Ref.~\cite{g3pi:radcorr} for the similar process $\gamma\pi^-\to\pi^-\pi^0$ that 
there is one significant effect due to radiative corrections, which is due to photon exchange in the $t$-channel,
compare Fig.~\ref{fig:photon-exchange}.
Translated to the process under investigation here,
the inclusion of this effect amounts to correcting the scattering amplitude in the form
\beq
\F(s,t,u) \to \F(s,t,u) - \frac{2e}{t}A^\eta_{\gamma\gamma}. \label{eq:photon}
\eeq
Strictly speaking, the photon-exchange amplitude would have to be amended by
form factor effects, including both the $\eta$ transition and the pion vector form factor; 
however, these corrections were shown to be very small in $\gamma\pi^-\to\pi^-\pi^0$~\cite{g3pi:radcorr}.
The inclusion of the correction~\eqref{eq:photon} may be desirable if experimental data on
$\gppe$ become sufficiently precise in the future; we still neglect it for the following investigation. 
\esp

\begin{figure}
\includegraphics[width=\linewidth]{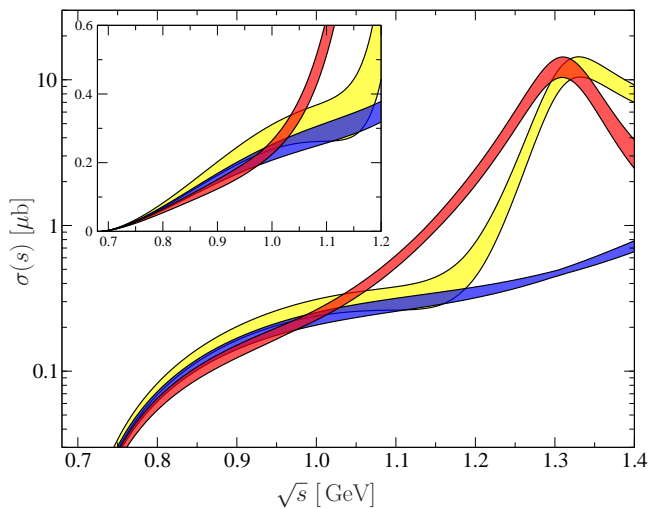}
\caption{Total cross section $\sigma(s)$ for $\gamma\pi^-\to\pi^-\eta$. 
The blue band shows the cross section obtained from crossing the decay amplitude
of Ref.~\cite{Stollenwerk}; the red band corresponds to the full amplitude including 
$a_2$ effects.
Finally, the yellow band displays the full cross section for the relative sign of the $a_2$ 
contribution flipped.
The insert magnifies the near-threshold region. See main text on the error bands.}\label{fig:totalX}
\end{figure}

We show the total cross section in Fig.~\ref{fig:totalX}.  
We compare the cross section obtained 
from the decay amplitude in Ref.~\cite{Stollenwerk} by crossing to the full cross section including 
$a_2$ effects.  We find that dominance of $t$-channel dynamics holds roughly up to $\sqrt{s} =1\GeV$, 
while above this value, the tensor resonance begins to dominate.  We predict a peak cross section of about 
$(12\pm2)\mu\text{b}$, which is of a similar order of magnitude as the cross section of 
$\gamma\pi^-\to\pi^-\pi^0$ at the $\rho$ peak~\cite{g3pi}.
\begin{figure*}
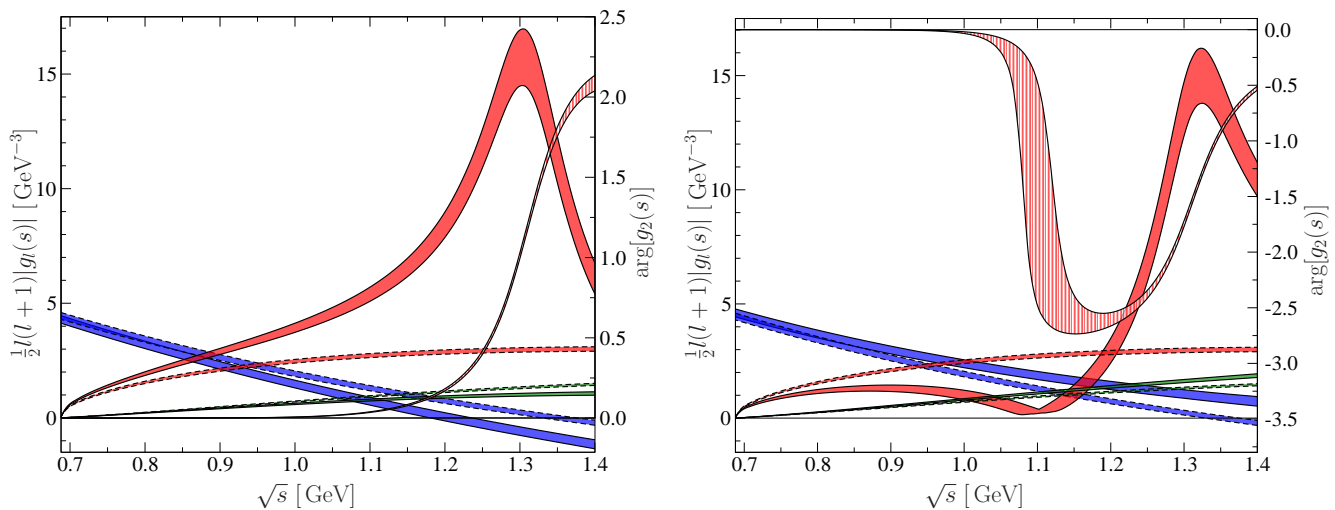

\includegraphics[width=0.49\linewidth]{PW.eps} \hfill
\includegraphics[width=0.49\linewidth]{PW_wrongsign.eps}
\caption{First three partial waves for $\gamma\pi^-\to\pi^-\eta$.  
The moduli are shown in the normalization $\frac{1}{2}l(l+1)|g_l(s)|$ for $P$-wave (blue bands), 
$D$-wave (red), and $F$-wave (green); bands with dashed borders refer to the analytic continuation
of the decay amplitude in Ref.~\cite{Stollenwerk}, while the full bands show the full result
including $a_2$ effects.  The phase of the complete $D$-wave is represented by the red-striped band.
All indicated bands combine the uncertainties in $\Gamma(\eppg)$, $\alpha_\Omega$, 
and the $a_2$ couplings $c_Tg_T$.
}\label{fig:PW}
\end{figure*}
For completeness, we also display the cross section with the relative sign of the $a_2$ contribution, 
see Eq.~\eqref{eq:sign-a2}, flipped (and all other parameters adjusted such as to best reproduce the 
$\eppg$ decay data); we see that the transition from the near-threshold to the 
resonance region looks quite different, for reasons that will become transparent below.
The uncertainty in the resonance peak is obviously dominated by those in the $a_2$ coupling constants $c_Tg_T$,
while near threshold, the errors coming from the total decay rate $\Gamma(\eppg)$ 
as well as $\alpha_\Omega$ are more important.

In the introduction, we pointed out that a naive continuation of the $\eppg$ decay amplitude
Eqs.~\eqref{eq:Omnes} and \eqref{eq:P} would lead to a zero in the scattering amplitude $\gamma\pi^-\to\pi^-\eta$
at $t = -1/\alpha_\Omega$.  As $s$ increases, this zero first appears in the differential cross section 
$\diff\Gamma/\diff z_s$ in backward direction, i.e.\ for $z_s=-1$.  
Given the form of the partial-wave expansion~\eqref{eq:s-pwa},
\beq
\F(s,t,u) = g_1(s) + 3z_s g_2(s) + \ldots,
\eeq
and assuming $F$- and higher partial waves are small, this will occur once the $D$-wave is one third the 
size of the $P$-wave, as long as relative phases are small.
In our amplitude representation, the only imaginary part stems from the energy-dependent width of 
$s$-channel $a_2$ exchange; the $P$-wave phase is neglected, and all partial waves induced by $t$-channel
exchange are obviously real.
For better comparison and due to
\beq
P'_l(-1) = \frac{(-1)^{l-1}}{2}l(l+1),
\eeq
we display the first three partial waves multiplied with $\frac{1}{2}l(l+1)$ in Fig.~\ref{fig:PW};
the intersection of $P$- and $D$-wave curves then gives an indication at the energy at which an 
additional zero in the angular distribution will occur, with the precise position slightly modified by the small,
but nonnegligible $F$-wave.  We compare the full amplitude including the $a_2$ to the continuation of the 
decay amplitude from Ref.~\cite{Stollenwerk}.  The decisive observation is that including the $a_2$, the 
$D$-wave becomes more important than the $P$-wave at even lower energies, around $\sqrt{s}=0.9\GeV$, where
the phase is still tiny---we therefore indeed expect to observe an almost perfect vanishing of the amplitude.
To demonstrate that this is not trivially so, Fig.~\ref{fig:PW} also shows 
what would happen with the opposite sign of the $a_2$ contribution: negative interference of $s$-channel $a_2$
and $t$-channel exchange leads to a near-vanishing of the $D$-wave around $1.1\GeV$ (which
is the cause for the rapid phase variation at that energy), and its rise toward the 
$a_2$ peak only overtakes the $P$-wave once the phase is significant.
As a consequence, no near-complete cancelation ever occurs at any energy.

We wish to re-emphasize that there is no fixed relation between the phase of our $s$-channel partial waves
to $\pi\eta$ scattering phase shifts according to a final-state theorem.
As the corresponding $\pi\eta$ phases are not theoretically determined in the way the $\pi\pi$~\cite{Madrid,ACGL,CCL}
or $\pi K$~\cite{Orsay} phases are, unitarization using model phases seems to offer no significant improvement.
Furthermore, the $a_2$ is a largely inelastic resonance with respect to $\pi\eta$ scattering anyway,
with the dominant decay channel being $\pi\rho$ [see Eq.~\eqref{eq:a2width}], such that no simple version of 
Watson's theorem applies, and any unitarization would have to implement a coupled-channel formalism.

\begin{figure*}
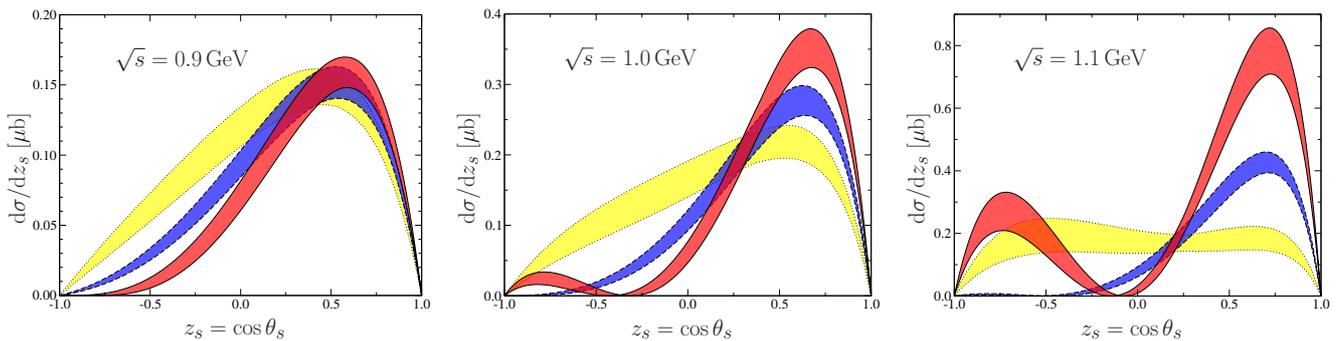

\includegraphics[width=0.32\linewidth]{dsig_0.9.eps} \hfill
\includegraphics[width=0.32\linewidth]{dsig_1.0.eps} \hfill
\includegraphics[width=0.32\linewidth]{dsig_1.1.eps} 
\caption{Differential cross sections $\diff\sigma/\diff z_s$ for the three energies 
$\sqrt{s} = 0.9\GeV$, $1.0\GeV$, and $1.1\GeV$ (from left to right). 
The blue bands denote analytic continuation of the amplitude of Ref.~\cite{Stollenwerk}, 
the red bands are our full predictions including $a_2$ effects, while the yellow bands
show the same with the opposite relative sign of the $a_2$ contributions.
The pronounced minima, very close to actual zeros, in backward direction in the red bands 
are clearly seen.}
\label{fig:dsig}
\end{figure*}
\bsp
For illustration, we also show the resulting angular distribution at three sample energies 
$\sqrt{s} = 0.9\GeV$, $1.0\GeV$, and $1.1\GeV$ in Fig.~\ref{fig:dsig}, 
indicating the transition between the threshold [$P$-wave dominance, $\diff\sigma/\diff z_s \propto (1-z_s^2)$]
and the resonance region [$D$-wave dominance, $\diff\sigma/\diff z_s \propto (1-z_s^2)z_s^2$].  
The very different features of the different signs of the $a_2$ are clearly visible.

\begin{figure}
\includegraphics[width=\linewidth]{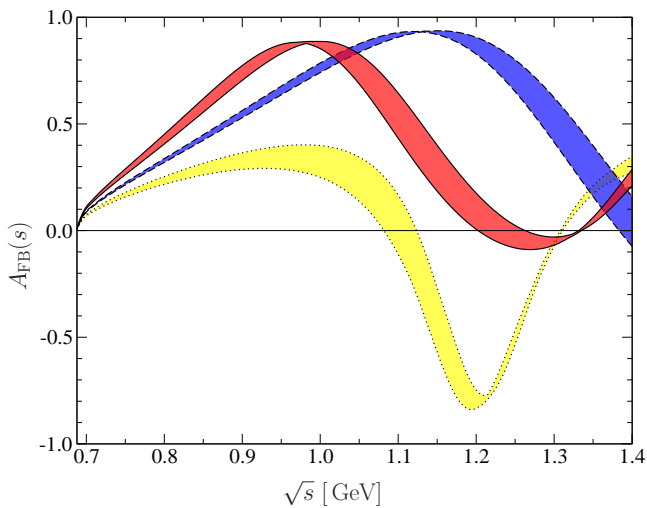} 
\caption{Forward--backward asymmetry according to Eq.~\eqref{eq:asym}. The color code is as in Fig.~\ref{fig:dsig}.}
\label{fig:asym}
\end{figure}
An observable that allows one to capture the key features of the effect
discussed even in a comparably low-statistics measurement
is to characterize the behavior
in terms of a forward--backward asymmetry,
\begin{align} \label{eq:asym}
A_{\text{FB}}(s) &= \sigma(s)^{-1} \bigg[
\int_0^1\diff z_s \frac{\diff\sigma}{\diff z_s} - \int_{-1}^0\diff z_s \frac{\diff\sigma}{\diff z_s}\bigg] \nl
&\simeq \frac{\Re(g_2)(g_1+g_3)}{\frac{4}{9}g_1^2+\frac{4}{5}|g_2|^2+\frac{8}{7}g_3^2},
\end{align}
where in the second line we have neglected all partial waves beyond $F$-waves, as well as
imaginary parts of the $P$- and the $F$-wave.  
As shown in Fig.~\ref{fig:asym}, both the continuation of the decay amplitude without $a_2$~\cite{Stollenwerk}
and our full model with the preferred relative sign for the $a_2$ display a very large positive 
asymmetry, peaked just below $\sqrt{s}=1\GeV$ for the full model; for the opposite $a_2$ sign,
the asymmetry is small near threshold, and subsequently even turns negative.
An experimental verification of this asymmetry would therefore confirm that our description 
of the decay amplitude including the $a_2$, and the resulting consequences for the $\eta$ transition form factor,
are indeed reasonable.
\esp

\section{Summary}\label{sec:summary}

In this article, we have studied the effects of the $a_2$ tensor meson on the decay $\eppg$
as well as the analytic continuation of the decay amplitude for the scattering process $\gppe$. 
We have included the $D$-wave $\pi\eta$ resonance as a left-hand cut structure of a dispersive
representation that obeys the correct final-state phase relation for the $\pi^+\pi^-$ $P$-wave.
While the decay spectra measured by the KLOE Collaboration can be described equally well
with and without the $a_2$ effects, there seems to be an indication for better consistency 
of the subtractions constants when comparing to the similar decay $\epppg$.
The slope parameter of the resulting $\eta$ transition form factor is reduced by about $7\%$ 
in the dispersive integral up to $1\GeV^2$ compared to a previous analysis~\cite{etaTFF}.  

We have predicted different observables for the $\eta$ production reaction $\gppe$ at energies
up to the $a_2$ resonance.  The peak cross section is predicted to be $(12\pm2)\mu\text{b}$, 
similar in size to the $\gamma\pi^-\to\pi^-\pi^0$ cross section in the $\rho$ peak~\cite{g3pi}.
Fixing the relative sign of the $a_2$ to the more likely solution
from decay phenomenology, we find an interesting $P$--$D$-wave interference effect, leading 
to almost perfect zeros in the differential cross section, and a very strong forward--backward asymmetry
in the energy region between threshold and the $a_2$ peak.
These predictions provide strong motivation to study the corresponding
Primakoff reaction e.g.\ at COMPASS,  which may help to further scrutinize 
the physics of light mesons relevant for hadronic corrections
to the muon's anomalous magnetic moment.

\bsp
\begin{acknowledgements}
We would like to thank Christoph Hanhart, Martin Hoferichter, 
and Andreas Wirzba for useful discussions and comments on the manuscript, 
and Andrzej Kup\'s\'c for 
supplying us with the acceptance-corrected data from Ref.~\cite{KLOE}.
Financial support by
the DFG (SFB/TR 16, ``Subnuclear Structure of Matter'')
and the Bonn--Cologne Graduate School
of Physics and Astronomy is gratefully acknowledged.
\end{acknowledgements}
\esp

\appendix

\section{\boldmath{$\eta'\to\pi\pi\gamma$}}\label{app:etaprime}

\bsp
The formalism of dispersively analyzing decay data with a final-state $P$-wave pion pair was
also applied to $\epppg$ in Ref.~\cite{Stollenwerk}, analyzing data by the
Crystal Barrel Collaboration~\cite{CB:etaprime}.
We do not intend to make a prediction for the crossed Primakoff reaction $\gamma\pi\to\pi\eta'$---the
threshold is too high, too close to the $a_2$ resonance tail to still find traces of the $t$-channel
exchange, and the number of inelastic (subthreshold) channels probably too large to be ignored.
However, we expect the impact of left-hand cuts in the decay process 
to be much stronger over the wider kinematic range accessible in the $\eta'$ decay, i.e., 
the curvature effects that are rather moderate in the $\eta$ decay in Fig.~\ref{fig:KLOElinear} 
should be much more visible in that case.  
Furthermore, the decay $\epppg$ is about to be remeasured with increased precision by 
BESIII (see~\cite{BES:etaprime} for spectra not yet corrected for acceptance), 
such that a prediction for the curvature of the spectrum (after dividing out the universal
Omn\`es factor) ought to be very timely.
\esp

To determine the $a_2$ contribution to $\epppg$, we first need to fix the $a_2\to\pi\eta'$ 
coupling constant.  We can first do this by naively defining a coupling $g'_T$ without a Lagrangian, just through
the analogous relation to Eq.~\eqref{eq:TPPwidth}
\beq
\Gamma(a_2\to\pi\eta') = \frac{(g'_T)^2}{180\pi F_\pi^4}\frac{\lambda^{5/2}(\ma^2,\mpi^2,\mep^2)}{\ma^7},
\eeq
which, with $\BR(a_2\to\pi\eta') = (5.3\pm0.9)\times10^{-3}$, yields
\beq
|g'_T| = (25.5\pm2.3)\MeV . \label{eq:gPT}
\eeq
If we attempt to explain this value based on the single Lagrangian term~\eqref{eq:Lagr1}, we first need to 
amend the pseudoscalar field $\phi$ by the SU(3) singlet $\eta_0$,
\beq
\phi = \left( \begin{array}{cc} \pi^0 + \frac{\eta_8+\sqrt{2}\eta_0}{\sqrt{3}} & \sqrt{2}\pi^+ \\ \sqrt{2}\pi^- & - \pi^0+\frac{\eta_8+\sqrt{2}\eta_0}{\sqrt{3}} \end{array}\right) + \ldots .
\eeq
In Eq.~\eqref{eq:uphi}, we have simply identified the $\eta$ with the octet field $\eta_8$; 
if now we assume a simple, single-angle $\eta\eta'$ mixing scheme,
\begin{align}
|\eta\rangle &= \cos\theta|\eta_8\rangle -\sin\theta |\eta_0\rangle , \nl
|\eta'\rangle &= \sin\theta|\eta_8\rangle +\cos\theta |\eta_0\rangle ,
\label{eq:mixing}
\end{align}
we can explain the ratio of the couplings by a mixing angle of $\theta = (-12.4 \pm 2.7)^\circ$,
which is somewhat smaller than the standard value $\theta \approx -20^\circ$, but close enough that 
we are confident the difference can be explained by higher-order terms.  In particular, we can safely conclude
that the sign of $g'_T$ in Eq.~\eqref{eq:gPT} agrees with the one of $g_T$.\footnote{As a side remark, 
we point out that fixing an effective $a_2\to\pi\rho$ coupling constant from the known branching fraction 
$\BR(a_2\to\pi\rho)$ should also allow us to include $a_2$ effects  in 
the decays $\eta'\to4\pi$~\cite{eta4pi,BES:eta4pi}, thus going further beyond vector-meson dominance.}

\begin{figure}
\includegraphics[width=\linewidth,clip]{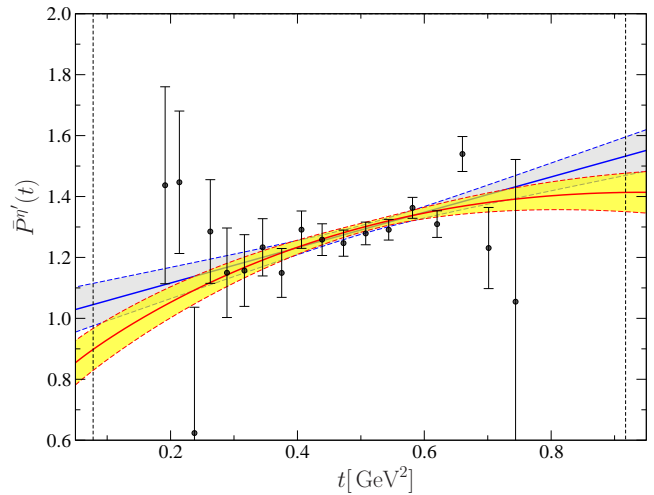}
\caption{Representation of the decay distribution $\epppg$ from Ref.~\cite{CB:etaprime};
see main text for details.  The blue curve shows the linear fit, including the gray band for the fit uncertainty.
The red curve with the yellow band
includes the effects of $a_2$ exchange in addition.  The vertical dashed lines represent
the limits of phase space at $4\mpi^2$ and $\mep^2$.} \label{fig:CBlinear}
\end{figure}

In Fig.~\ref{fig:CBlinear}, we display the observable $\bar P^{\eta'}(t)$ defined in strict analogy to 
Eq.~\eqref{eq:defPbar}, comparing the data of Ref.~\cite{CB:etaprime} to fits with a linear parametrization,
as well as including effects of the $a_2$.  Due to the rather large error bars, we show the fit results as bands,
not just the best fit.  The linear fit leads to a slope parameter 
\beq \label{eq:alphaprime-fit1}
\alpha_\Omega' = (0.6\pm0.2)\GeV^{-2},
\eeq
with a reduced $\chi^2$ of $\chi^2/\text{ndof} = 1.23$.
It was argued in Ref.~\cite{Stollenwerk} that in the limit of a large number of colors, $\alpha_\Omega=\alpha_\Omega'$ should
be expected, so the slopes of the polynomial would agree for $\eta$ and $\eta'$ decay.  Comparing 
Eqs.~\eqref{eq:alpha-fit1} and \eqref{eq:alphaprime-fit1}, phenomenology seems rather at odds with this prediction.
However, including the effects of the $a_2$ in the amplitude representation, we find much stronger curvature
effects than for the $\eta$ decay as anticipated, with the residual slope fitted to be
\beq \label{eq:alphaprime-fit2}
\alpha_\Omega' = (1.4\pm0.4)\GeV^{-2},
\eeq
now with $\chi^2/\text{ndof} = 1.38$. 
The additional uncertainty due to the $a_2$ couplings is $\pm 0.1\GeV^{-2}$.
In this case, the fit quality becomes slightly worse
(overall better fits are essentially precluded by the third-to-last data point at $\sqrt{t} = [800,825]\MeV$); 
however, $\alpha_\Omega'$ is now in markedly better agreement with the value found for $\alpha_\Omega$ in Eq.~\eqref{eq:alpha-fit2}.
$\bar P^{\eta'}(t)$ can be approximated in the decay region $4\mpi^2\leq t \leq \mep^2$ by a quadratic polynomial
\beq
\bar P^{\eta'}(t) \approx P^{\eta'}(0) \Big( 1+\bar \alpha'_\Omega t+ \bar \beta'_\Omega t^2 \Big)
\eeq
to about 1\% accuracy.  The $a_2$ contribution predicts the curvature to be 
$\bar\beta'_\Omega = (-1.0\pm0.1)\GeV^{-4}$.
As a side remark, we can also take this result as another strong indication on the correctness of the sign of $c_Tg_T$
and $c_Tg'_T$: a negative sign would lead to a residual slope $\alpha_\Omega'$ of $(0.06\pm0.12)\GeV^{-2}$.

A more rigorous test of the decay spectrum predicted here, with higher-statistics data from BESIII, would be 
extremely welcome.

\end{document}